\documentclass[twocolumn,secnumarabic,amssymb,nobibnotes,aps,showpacs,nofootinbib,noeprint]{revtex4-1}
\usepackage[english]{babel}
\usepackage{graphicx}

\usepackage{textcomp}

\usepackage{amsmath}
\usepackage{amsfonts}
\usepackage[usenames,dvipsnames]{color}
\usepackage{units}
\usepackage{subcaption}

\usepackage{hyperref}
\usepackage{rotating}
\usepackage{booktabs} 
\usepackage[sort&compress]{cleveref}
\usepackage{slashed}
\usepackage{transparent}
\usepackage{epstopdf}

\newcommand{\kB}{k_\textrm{B}}

\newcommand{\mhc}{\mathcal}

\newcommand{\lla}{\left\langle}
\newcommand{\rra}{\right\rangle}

\DeclareMathAlphabet{\mathbb}{U}{bbold}{m}{n}

\begin{document}
\title{Evaluation of memory effects at phase transitions and during relaxation processes}

\author{Hugues Meyer}
\affiliation{Department of Theoretical Physics \& Center for Biophysics, Saarland University, 66123 Saarbrücken, Germany}
\author{Fabian Glatzel}
\affiliation{Institute of Physics, University of Freiburg, Hermann-Herder-Str. 3, 79104 Freiburg, Germany}
\author{Wilkin W{\"oh}ler}
\affiliation{Institute of Physics, University of Freiburg, Hermann-Herder-Str. 3, 79104 Freiburg, Germany}
\author{Tanja Schilling}
\email{Tanja.Schilling@physik.uni-freiburg.de}
\affiliation{Institute of Physics, University of Freiburg, Hermann-Herder-Str. 3, 79104 Freiburg, Germany}

\date{\today}

\begin{abstract}
We propose to describe the dynamics of phase transitions in terms of a non-stationary Generalized Langevin Equation for the order parameter. By construction, this equation is non-local in time, i.e.~it involves memory effects whose intensity is governed by a memory kernel. In general, it is a hard task to determine the physical origin and the extent of the memory effects based on the underlying microscopic equations of motion. Therefore we propose to relate the extent of the memory kernel to quantities that are experimentally observed such as the induction time and the duration of the phase transformation process. Using a simple kinematic model, we show that the extent of the memory kernel is positively correlated with the duration of the transition and of the same order of magnitude, while the distribution of induction times does not have an effect. This observation is tested at the example of several model systems, for which we have run Molecular Dynamics simulations: a modified Potts-model, a dipole gas, an anharmonic spring in a bath and a nucleation problem. All these cases are shown to be consistent with the simple theoretical model.
\end{abstract}

\maketitle

\section{Introduction}
Phase transitions are characterized by order parameters, such as an averaged local density that distinguishes a gas from a liquid, a magnetization that distinguishes a paramagnetic phase from a ferromagnetic phase, or a degree of local positional order that distinguishes a crystal from a liquid. For complex systems out of thermal equilibrium, it is in general not possible to compute the evolution of such coarse-grained observables exactly based on the underlying microscopic dynamics. (This would require to integrate out a large number of degrees of freedom from a time-dependent distribution of micro-states, which is a hard problem.) Therefore, the dynamics of order parameters is often modelled by means of approximative theories\cite{Chaikin1995} such as phase field models\cite{steinbach2009}, dynamic density functional theory \cite{marconi:1999}, Cahn-Hilliard theory \cite{cahn:1961} or classical nucleation theory\cite{Volmer1926,Becker1935,Oxtoby1998}. In these approaches the equation of motion for the evolution of the order parameter is usually assumed to be time-local. However, if one derives the evolution equation of a coarse-grained observable from the underlying microscopic dynamics of a non-equilibrium system, in general one obtains an equation of motion with memory \cite{zwanzig:1961,mori:1965,grabert:1982,Meyer2019}. We therefore discuss the question: What is the temporal extent of memory during the evolution of an order parameter at a phase transition and under which conditions can it be neglected?

As the physical interpretation of memory effects -- and of their origin -- is in general not straightforward, we draw a connection between simple phenomenological observations of the dynamics of phase transitions and the quantification of memory kernels of an exact non-equilibrium coarse-grained model. We first present a theoretical estimate and then compare the estimate to the results of four computer simulations studies.

\section{Averages over bundles of trajectories}
\label{sec:theory}
We begin by recalling how an equation of motion for a coarse-grained observable can be derived.
Consider a system of a large number of microscopic degrees of freedom $\left\{\Gamma_{i}\right\}$, $i=1 \ldots N$, that evolve according to Hamilton's equations of motion, and a phase-space observable $A(\mathbf{\Gamma})$ (e.g.~an order parameter of a phase transition) which is a well defined function at any point in phase space $\mathbf{\Gamma} = (\Gamma_1, \ldots , \Gamma_N)$. Then prepare a non-equilibrium ensemble, {\it i.e.}~a large number of copies of this system initialized according to any phase-space distribution of interest.  
We showed in ref.~\cite{Meyer2017,Meyer2019} by means of a projection operator formalism, that for any such observable, any initial ensemble and any dynamical process one can define functions $\omega(t)$, $K(t',t)$ and $\eta_{t}$ such that the equation of motion for $A_{t}$ is
\begin{equation}
  \label{EOM_A}
  \frac{d A_{t}}{dt} = \omega (t) A_{t} + \int_{0}^{t} d\tau  K(\tau, t) A_{\tau} + \eta_{t} \quad ,
\end{equation}
where $K(\tau,t)$ is the memory kernel and the average is taken over the ensemble of non-equilibrium trajectories, i.e.~over the evolving distribution of microstates. 
The letter $t$ as a subscript denotes the time-dependence on a single trajectory, whereas the time-dependence between parentheses indicates a function of time independent of the trajectory, which is determined by the choice of initial ensemble. The form of the equation is thus a non-stationary version of the Generalized Langevin equation. Note that $\omega(t) = \text{d}\left( \ln \sqrt{C(t,t)} \right)/\text{d}t$.
A similar equation holds for the auto-correlation function $C(t',t) = \left\langle A^{*}(t') A(t) \right\rangle$
\begin{equation}
    \label{EOM_C}
    \frac{\partial C(t',t)}{\partial t} = \omega (t) C(t',t) + \int_{t'}^{t} d\tau C(t',\tau) K(\tau, t)
\end{equation}

In the case of phase transitions, the observable has a time-dependent average. For such observables equation~(\ref{EOM_A}) can be inconvenient, because the contributions of $\omega$ and $K$ to the average $\left\langle A(t) \right\rangle$ and to the fluctuations around the average are coupled. To resolve these contributions, we consider instead the equation of motion of a shifted and normalized quantity
		\[
                  \Delta\tilde{A}_{t} := \frac{ A_{t} - \left\langle A(t) \right\rangle }{\sqrt{\left\langle A(t)^{2} \right\rangle - \left\langle A(t) \right\rangle^{2}}}\quad .\]
                $\Delta\tilde{A}_{t}$ fullfills an equation of the same form as equation(\ref{EOM_A}) with a memory kernel $\Delta\tilde{K}(\tau,t)$ and a flucutaing term $\Delta\tilde{\eta}_{t}$.
		The equation of motion for $A_{t}$ can then be rewritten as
		\begin{eqnarray}
		\label{shifted_nsGLE}
                  \frac{dA_{t}}{dt} &=& \frac{d}{dt} \left\langle A(t) \right\rangle +  \Delta \omega(t) \Delta A_{t}  \\
                  &&+ \nonumber \int_{0}^{t} d\tau \Delta\tilde{K}(\tau,t) w(\tau,t) \Delta A_{\tau} + \left\langle \Delta A(t)^{2} \right\rangle \Delta\tilde{\eta}_{t}
		\end{eqnarray}
		with $\Delta A_{t} = A_{t} - \left\langle A(t) \right\rangle$, $w(t',t) =  \sqrt{\left\langle \Delta A(t)^{2} \right\rangle/\left\langle \Delta A(t')^{2} \right\rangle}$ and $\Delta \omega(t) = \frac{d}{dt}\ln \sqrt{\left\langle \Delta A(t)^{2} \right\rangle}$. The first term on the right side of Equation~(\ref{shifted_nsGLE}) describes the evolution of the mean of the observable, the second term describes how much a single trajectory deviates from the mean trajectory.

Now let us assume we had observed the dynamics of a phase transition in an experiment or a simulation by recording an order parameter $A$, the value of which is close to zero in the initial phase, and $\simeq 1$ in the final phase. Phenomenologically, the trajectory $A_{t}$ will often be of the form 
\begin{equation}
  \label{eq:OPdynamics}
A_{t} = \frac{1}{1 + e^{-\frac{t-T}{\Delta}}} + \xi_{t} \; ,
\end{equation}
where $T$ is a random time, that corresponds to the time at which the transition occurs -- in the case of a first order transition, this would be the induction time, while in the case of a critical transition, $T$ would be close to zero on all trajectories. Here, we will not yet impose any conditions on the distributions of $T$, but just assume that for each trajectory, $T$ is drawn from a distribution $p_{T}(T)$ which characterizes a specific transition. $\Delta$ is the timescale needed by the system to undergo the transition, we will therefore refer to it as the {\it duration of the transition}. We set it to a constant value in this model, but we should note that its distribution $p_{\Delta}$ in realistic systems is actually not infinitely narrow. Our model will be consistent  as long as the width of $p_{\Delta}$ is much smaller than the one of $p_{T}$. Finally, $\xi_{t}$ are the fluctuations around the transition, whose amplitude we assume to be small and whose average $\lla \xi(t) \rra = 0$ we assume to vanish.

\subsection{The case $\Delta \to 0$} We first consider the unrealistic, but analytically tractable limit case of a discontinuous transition, which corresponds to $\Delta \to 0$ and
\begin{equation}
A_{t} = \left\{
\begin{tabular}{ll}
$\xi_{t}$ & if $t\leq T$ \\
$1+\xi_{t}$ & if $t > T$
\end{tabular}
\right.
\end{equation}
The average $\lla A(t) \rra$ can be computed in terms of the probability distribution $p_{T}(T)$ as 
\begin{equation}
\lla A(t) \rra = \int_{0}^{t} d\tau p_{T}(\tau) 
\end{equation}
Here we have used the fact that the average of the noise $\xi$ vanishes, and that the main contribution of a trajectory to the average is $0$ if $t\leq T$. Similarly, for each trajectory, the quantity $(A_{t}-\xi_{t})(A_{t'}-\xi_{t'})$ is equal to $1$ if both $t$ and $t'$ are greater than $T$, but vanishes otherwise. We thus have 
\begin{equation}
\left\langle A(t')A(t) \right\rangle = \left\{
\begin{tabular}{ll}
$\int_{0}^{t'} d\tau \ p_{T}(\tau) + \lla \xi(t')\xi(t)\rra$ & if $t'\leq t$ \\
$\int_{0}^{t} d\tau \ p_{T}(\tau) + \lla \xi(t')\xi(t)\rra$ & if $t\leq t'$
\end{tabular}
\right.
\end{equation}
We show in Appendix \ref{xit} that the overall effect of the noise term $\lla \xi(t')\xi(t)\rra$ on the memory kernel results in a time-local contribution in the form of a Dirac delta-distribution and a global rescaling prefactor close to $1$ for $|t-t'|>0$. We thus discard the autocorrelation function of $\xi$ in the next lines, because we intend to focus on non-local contributions and the global prefactor will not impact the evaluation of the timescale of the memory effects. We consider 
\begin{equation}
\left\langle A(t')A(t) \right\rangle = \left\{
\begin{tabular}{ll}
$\int_{0}^{t'} d\tau \ p_{T}(\tau)$ & if $t'\leq t$ \\
$\int_{0}^{t} d\tau \ p_{T}(\tau)$ & if $t\leq t'$
\end{tabular}
\right.
\end{equation}
To simplify the expressions in the next lines we define
\begin{align}
\left\{
\begin{tabular}{l}
$f_{T}(t) := \int_{0}^{t} d\tau \ p_{T}(\tau)$  \\
$h_{T}(t) := f_{T}(t)\left(1-f_{T}(t)\right)$  
\end{tabular}
\right.
\end{align}

The shifted and modified time auto-correlation function $\Delta\tilde{C}(t',t) = \lla \Delta\tilde{A}(t')\Delta\tilde{A}(t) \rra$ evolves as 
\begin{equation}
\Delta\tilde{C}(t',t) = \left\{
\begin{tabular}{ll}
$\frac{f_{T}(t')\left(1 - f_{T}(t)\right)}{\sqrt{h_{T}(t')h_{T}(t)}}$ & if $t'\leq t$ \\
$\frac{f_{T}(t)\left(1 - f_{T}(t')\right)}{\sqrt{h_{T}(t')h_{T}(t)}}$ & if $t\leq t'$
\end{tabular}
\right.
\end{equation}
We note that for any triplet of times $t'<s<t$
\begin{align}
\Delta\tilde{C}(t',s)\Delta\tilde{C}(s,t) =& \frac{f_{T}(t')\left(1 - f_{T}(s)\right)}{\sqrt{h_{T}(t')h_{T}(s)}}\frac{f_{T}(s)\left(1 - f_{T}(t)\right)}{\sqrt{h_{T}(s)h_{T}(t)}} \\ 
=& \frac{f_{T}(t')\left(1 - f_{T}(t)\right)}{\sqrt{h_{T}(t')h_{T}(t)}} \\
=& \Delta\tilde{C}(t',t)
\end{align}
This implies that $\Delta\tilde{C}(t',t) = \exp\left(\int_{t'}^{t} d\tau \gamma(\tau)\right)$, and that the memory kernel is of the form $\Delta\tilde{K}(t',t) \propto \delta(t-t')$, i.e.~there is no memory and the evolution equation of the order parameter is local in time. More precisely, we show in Appendix \ref{Delta0} that
\begin{equation}
	\Delta\tilde{K}(t',t) = -\frac{p_{T}(t)}{h_{T}(t)} \delta(t'-t)
\end{equation}

\subsection{The case $\Delta > 0$}
Let us now turn to the general case, in which the transition from $0$ to $1$ is not instantaneous but takes a non-zero duration of time $\Delta >0$. An analytic derivation of the memory kernel is no longer possible here, therefore we use a numerical method that we introduced in ref.~\cite{Meyer2020}.

\begin{figure}
	\begin{center}
		\includegraphics[width=0.49\linewidth]{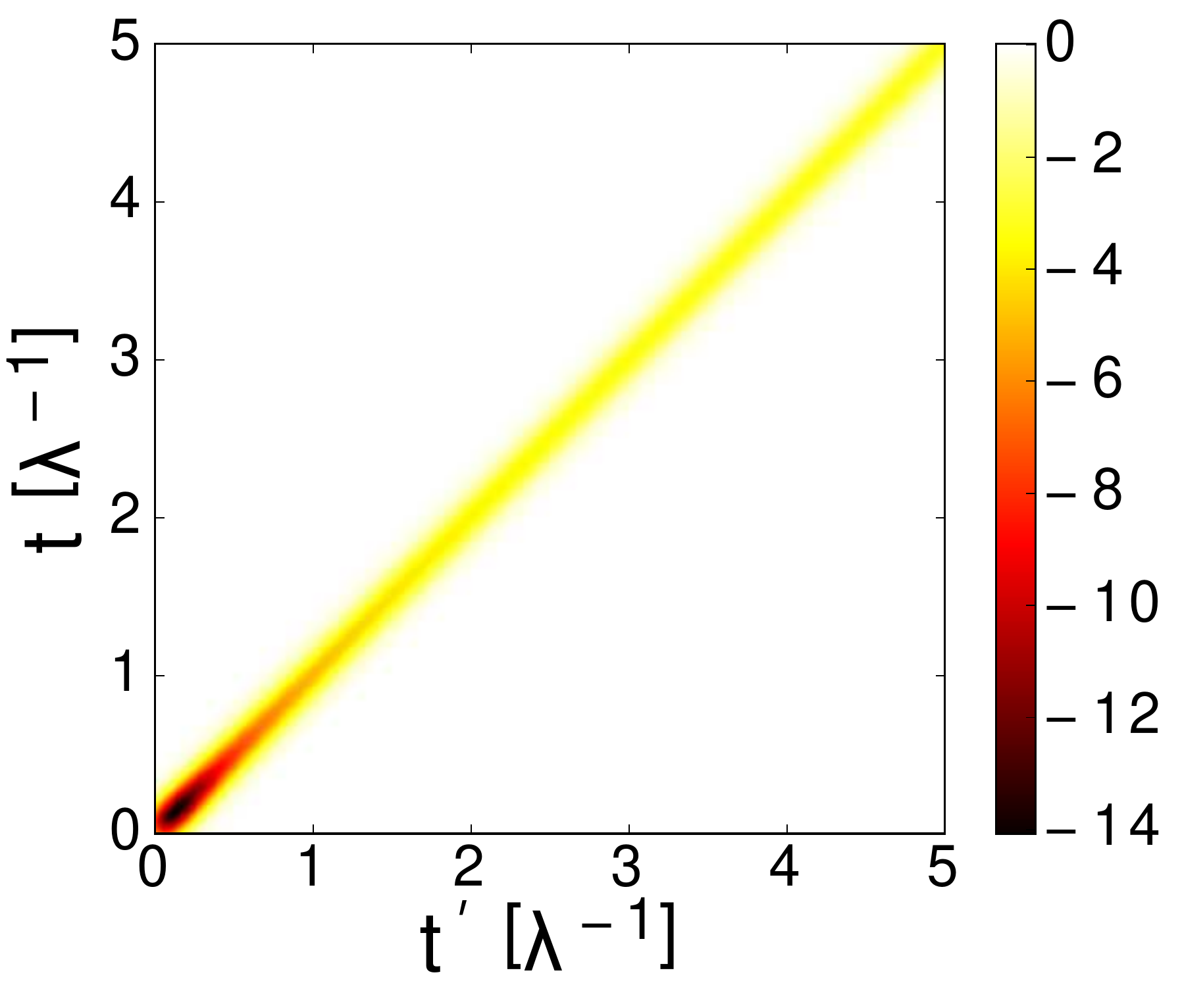}
		\includegraphics[width=0.49\linewidth]{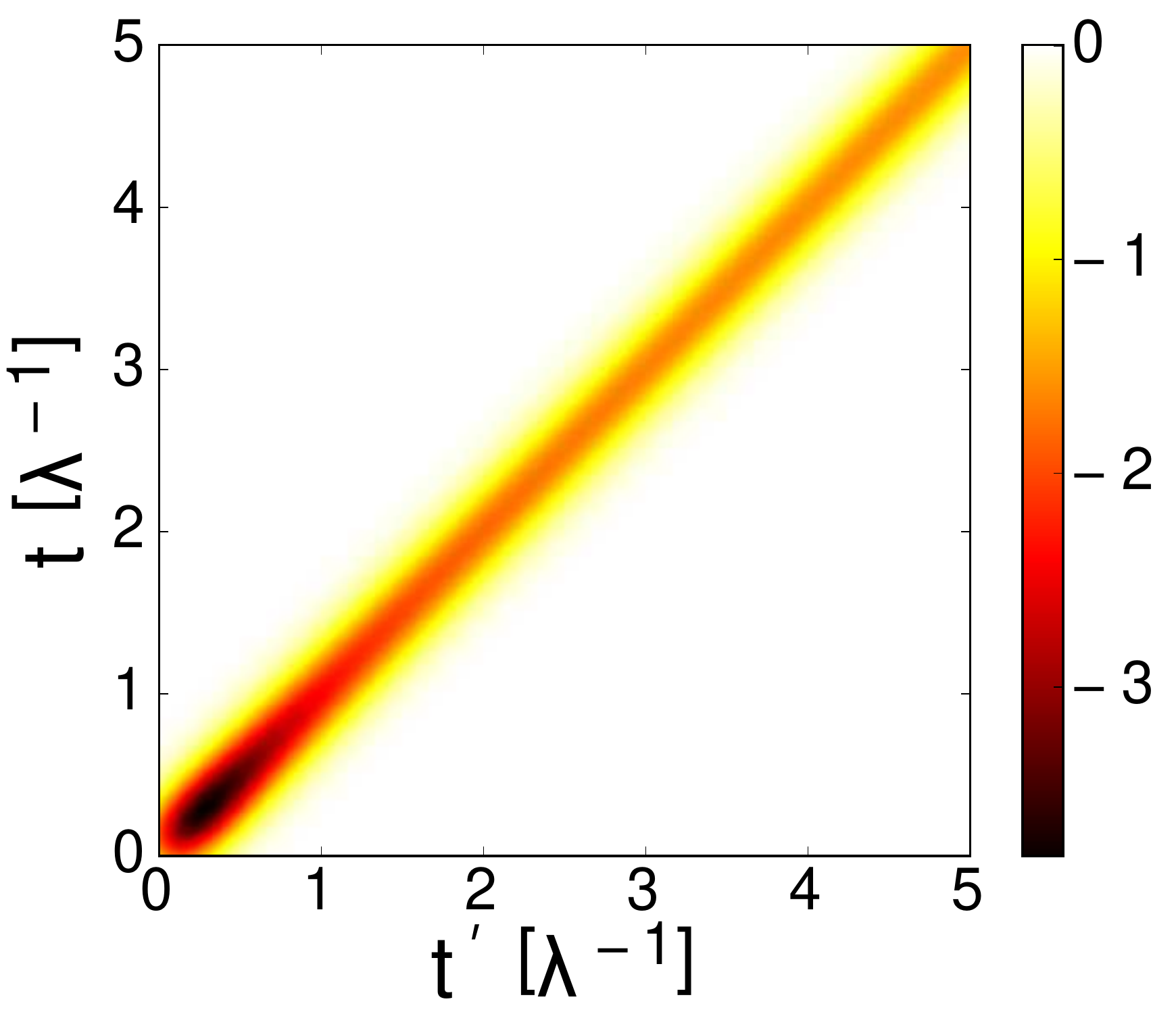}
		\includegraphics[width=0.49\linewidth]{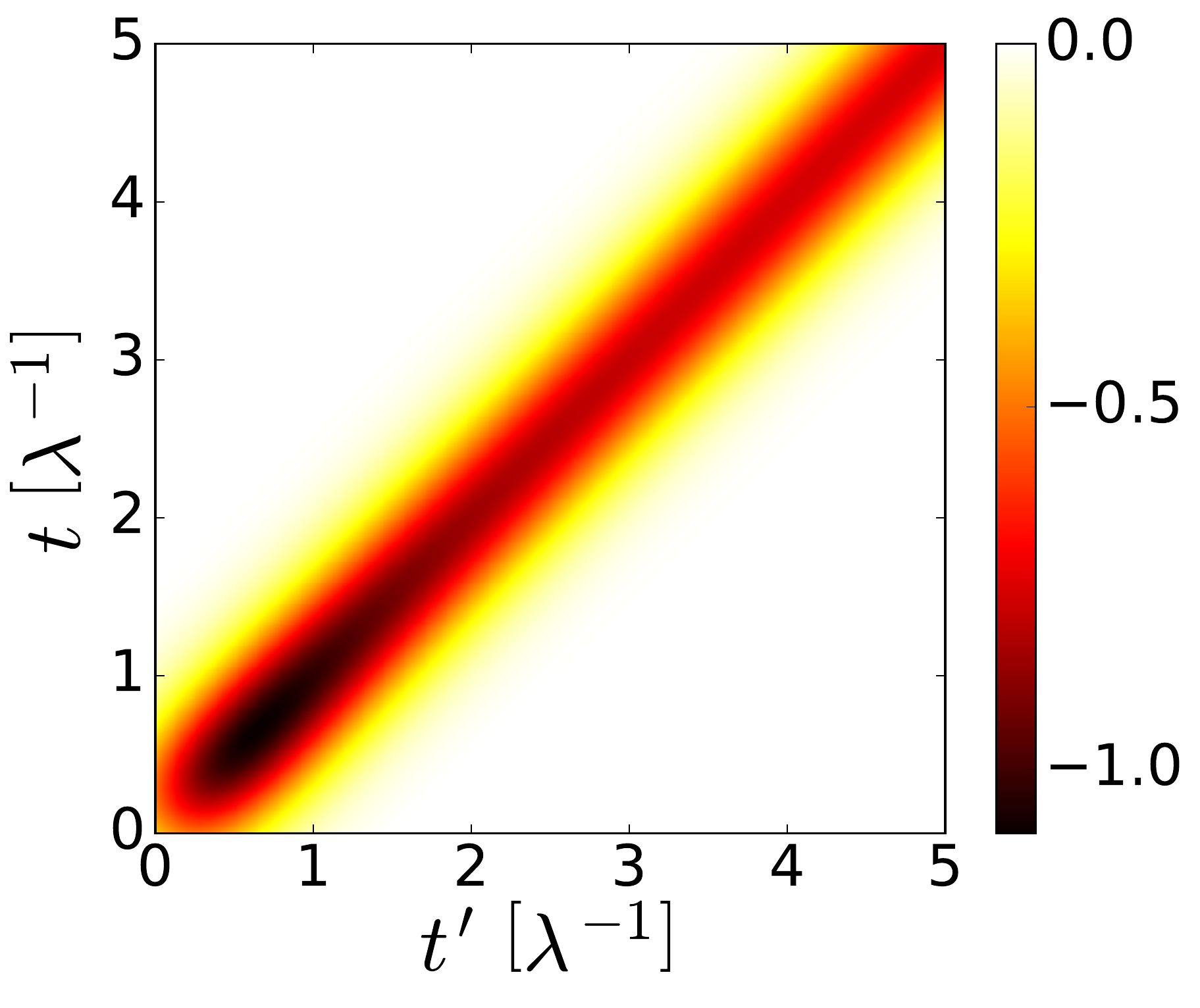}
		\includegraphics[width=0.49\linewidth]{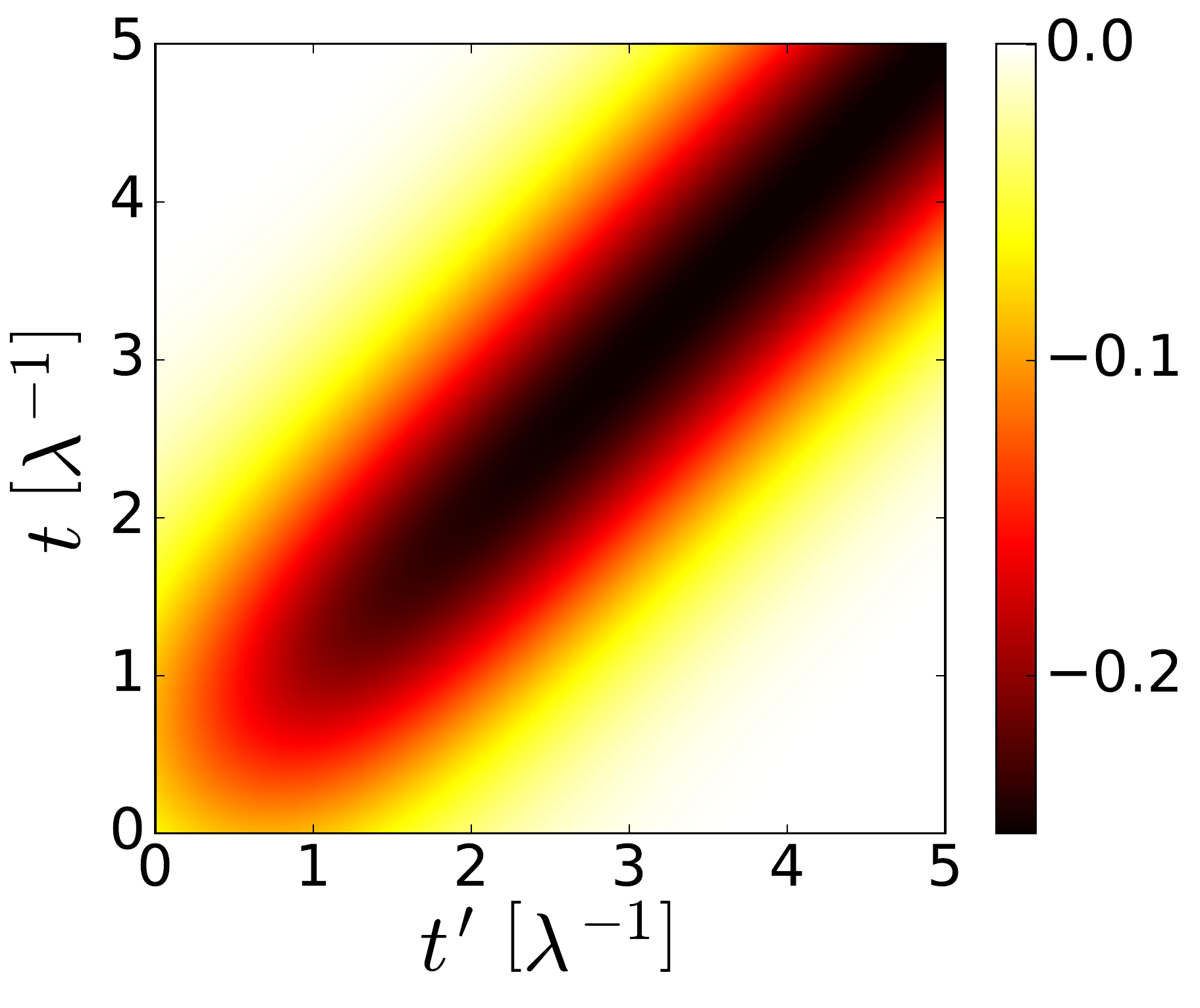}
	\end{center}
	\caption{Memory kernel $\Delta\tilde{K}(t',t)$ (in units of $\lambda^{-2}$) as a function of both $t$ and $t'$ for $\Delta = 0.05\lambda, 0.1, 0.2, 0.5$ (in units of $\lambda^{-1}$, from top to bottom, left to right). Brightness (colour online) indicates the value of $\Delta\tilde{K}(t',t)$. {\bf TS: something has gone wrong with the units in the caption. Please cross-check}}
	\label{kernels_poisson}
\end{figure}

\begin{figure}
	\begin{center}
		\includegraphics[width=0.99\linewidth]{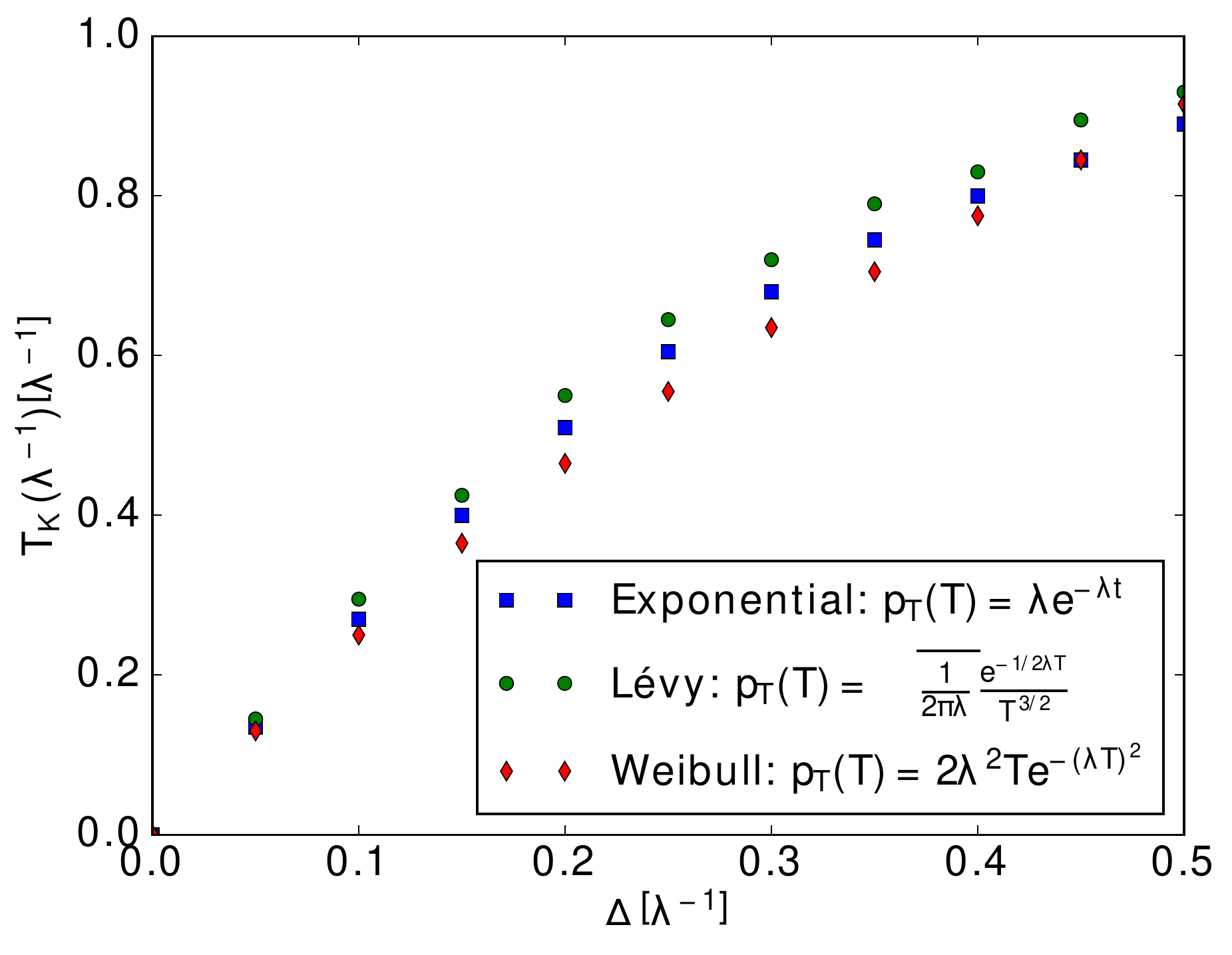}
	\end{center}
	\caption{Time extent of the memory kernel as a function of the duration of the transition $\Delta$ for three passage time distributions $p_{T}(T)$. The extent of the kernel and $\Delta$ are of the same order of magnitude.}
	\label{widths_kernels}
\end{figure}

Given a probability distribution $p_{T}(T)$ we compute the average $\lla A(t) \rra$ and correlation function $\lla A(t') A(t) \rra$  
\begin{align}
\lla A(t) \rra =& \int_{0}^{\infty} \frac{p_{T}(\tau) d\tau}{1 + e^{-\frac{t-\tau}{\Delta}}} \\
\lla A(t') A(t) \rra =& \int_{0}^{\infty} \frac{p_{T}(\tau) d\tau}{\left(1 + e^{-\frac{t'-\tau}{\Delta}}\right)\left(1 + e^{-\frac{t-\tau}{\Delta}}\right)} 
\end{align}
Again, the autocorrelation of the noise $\lla \xi(t)\xi(t') \rra$ is discarded here because its overall contribution to the timescale of the memory kernel is negligible (see Appendix \ref{xit}). We then define the shifted and normalized auto-correlation function $\Delta\tilde{C}(t',t)$ as before and numerically compute the corresponding memory kernel $\Delta\tilde{K}(t',t)$. We analyse three types of distributions: an exponential distribution $p_{T}(T)=\lambda e^{-\lambda T}$, a L\'{e}vy distribution $p_{T}(T) = \frac{1}{\sqrt{2\pi\lambda}} e^{-\frac{1}{2\lambda T}}T^{-3/2}$, and a Weibull distribution $p_{T}(T) = 2\lambda^{2} T e^{-(\lambda T)^{2}}$. For all of these distributions, we use $\lambda^{-1}$ as the unit of time. We show $\Delta\tilde{K}$ for the exponential distribution and various values of $\Delta$ in fig.~\ref{kernels_poisson}. The extent (``width'') of the memory kernel increases with increasing values of $\Delta$. In order to quantify this observation, we locate the time $t'<\lambda^{-1}$ for which $\Delta\tilde{K}(t',t) = \Delta\tilde{K}(t,t)/2$ at $t=\lambda^{-1}$. We call this time, which approximates the extent of the memory kernel, $T_K$. In fig.~\ref{widths_kernels} we show $T_K$ as a function of $\Delta$ for the three tested distributions. We conclude that the extent of the memory is determined mostly by the transition timescale $\Delta$, i.e.~the speed with which the phase transformation happens, rather than the details of the passage time distribution: the slower the transition, the longer the memory. As the data in fig.~\ref{widths_kernels} lies close to the bisecting line, we note that the extent of the memory kernel is of the same order of magnitude as the duration of the transition.

This simple semi-analytic approach indicates that the the duration of the transition $\Delta$ is strongly correlated with the extent of the memory kernel, while the distribution of induction times is irrelevant. However, the model used here is very simplistic. We thus need to corroborate the claim by data from more complex processes, in which the sigmoidal form for the evolution of the order parameters is not necessarily observed. We will therefore measure the memory kernels for four different Molecular Dynamics simulation data sets.

\section{Modified Potts-Model}
\begin{figure}
	\centering
	\includegraphics[width=0.99\linewidth]{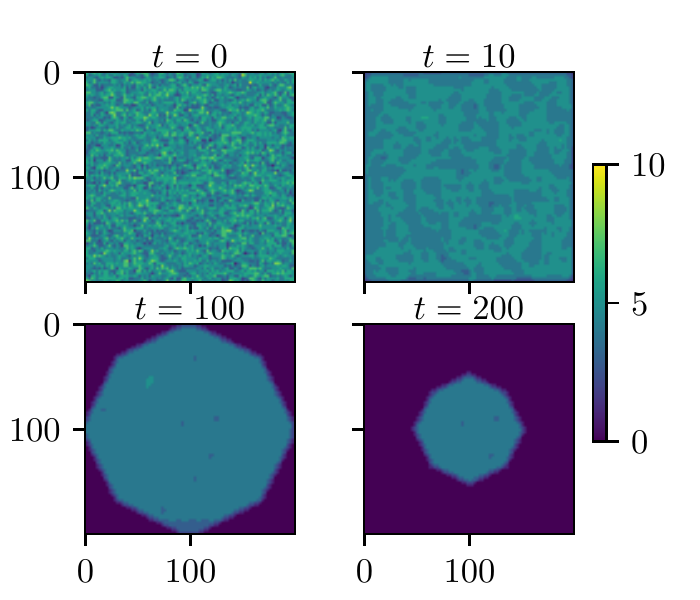}
	\caption{Exemplary snapshots of relaxation process of the modified Potts-model for $a=3$. Brightness (colour online) indicates spin values. }
	\label{fig:snapshots}
\end{figure}
\begin{figure}
	\centering
	\includegraphics[width=0.99\linewidth]{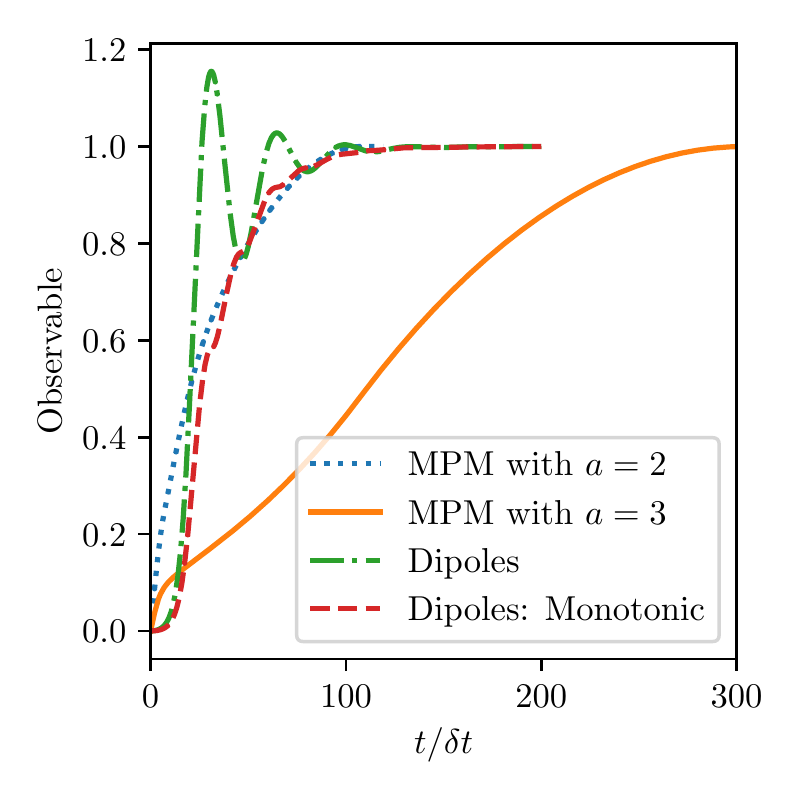}
	\caption{Scaled and shifted observables during relaxation processes: Average magnetization in the Potts model (dotted and solid line), averaged dipole moment along z-axis in the dipole model (dashed and dash-dotted line).}
	\label{fig:observables}
\end{figure}

As a first example model we study a two-dimensional modified Potts-model (MPM) with deterministic dynamics. The idea is to generate a simple relaxation process. The system is composed of spins on a $N\times N$ square lattice, each of which having $M$ possible states. Each spin has 8 neighbours (those with contact at the corners included). At $t=0$ the system is initialized by assigning all spins random values drawn from a uniform distribution. The system is then let free to evolve according to the following deterministic update rule: 
\begin{align}
	s_i(t+\delta t) &= \left\lfloor\frac{1}{8}\left(a + \sum\limits_{j}s_j(t)\right)\right\rfloor
\end{align}
Here, the sum runs over all neighbours of $s_i$ and $a$ is a bias. This rule ensures the spins to always be in one of the $M$ states. At the boundaries of the lattice, missing neighbours are treated as spins in the $0$-state, i.e. $s_j=0$. This ensures that the system becomes completely ordered in the long-time limit. We choose $\delta t=1$ as the unit of time because the evolution of the system is only defined on discrete time-steps.

As the order parameter we compute a scaled and shifted average off all spins: 
\begin{equation}
S(t)=c_1\sum s_i(t)+c_2.
\end{equation}
Here, $c_1$ and $c_2$ are chosen such that the averages over trajectories $\langle S(0)\rangle=0$ and $\lim\limits_{t\to\infty}\langle S(t)\rangle = 1$. On the level of individual trajectories, we get a sharp distribution of the observable around $S=0$ with a standard deviation of $\sigma_S=0.003$ at $t=0$. Two parameters sets are tested: $\left\{M=11, N=200, a=2 \right\}$ and $\left\{M=11, N=200, a=3 \right\}$, and at least 1000 trajectories are sampled for each of them. Exemplary snapshots of a system with $a=3$ are shown in \cref{fig:snapshots}. The time-evolution of the order parameter is shown in \cref{fig:observables} (MPM, dotted and solid line). 

\begin{figure}
	\centering
	\includegraphics[width=0.99\linewidth]{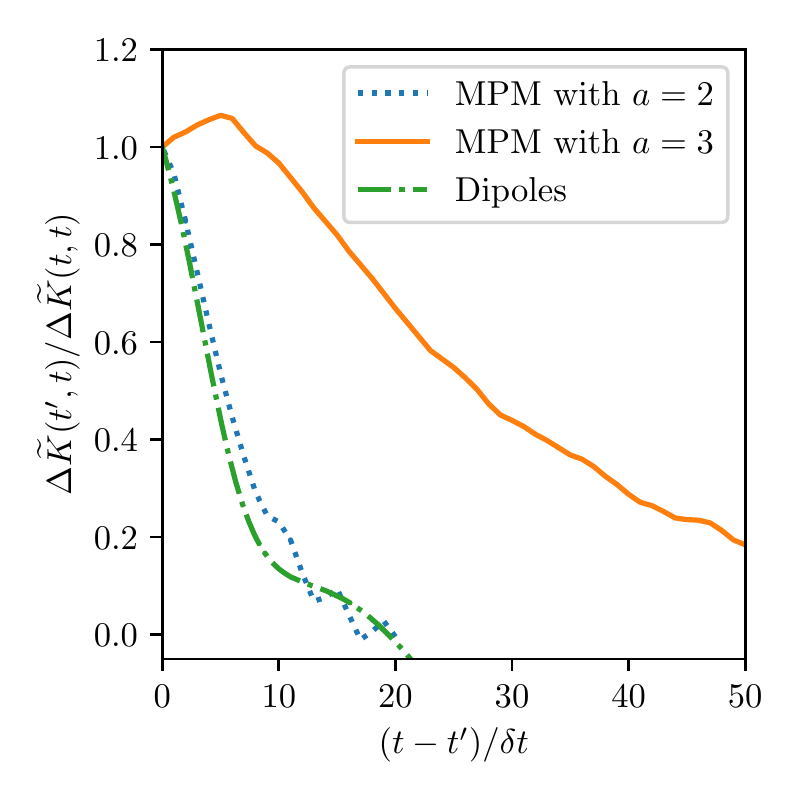}
	\caption{Kernels of different relaxation processes for times $t$ where the averaged observable is $0.5$. Potts model (dashed and dotted line) and dipole model (dash-dotted line).}
	\label{fig:kernelswidthcomparison}
\end{figure}

Since the shape of the order parameter curve deviates noticeably from the sigmoid function used before, we evaluate the extent of the kernel at the time where the observable averaged over all trajectories is $0.5$. (Note that the distribution of the times where the transitions occur is very sharp here. Thus, calculating the time where the transitions occur for every trajectory individually and averaging afterwards yields a result that deviates from the former value by no more than a single time step.) Then, the transition time is obtained by taking half of the time difference between the points where the observable intersects $1/(1+e)\simeq 0.27$ and $1/(1+e^{-1})\simeq 0.73$.  The kernels evaluated at those times are depicted in \cref{fig:kernelswidthcomparison}.

The ratios of the extents of the kernels and the durations of the transition are given in \cref{table:ratiosWidthTransitionTime}. The kernel extent and the transition duration are of similar magnitude, as predicted in section \ref{sec:theory}.
\begin{table}[]
	\begin{tabular}{c|c}
		Process & Ratio \\
		\hline
		\hline
		MPM with $a=2$&$0.33$\\
		MPM with $a=3$&$0.54$\\
		Dipoles&$0.45$\\
		Nucleation&$1.8$
	\end{tabular}
	\caption{Ratio of kernel extent to duration of transition for different types of relaxation processes and phase transitions.}
	\label{table:ratiosWidthTransitionTime}
\end{table}

\section{Dipole Gas}
As a second example we consider the relaxation process in a gas of dipoles, in which the particles align in the direction of an external homogeneous field. Here, 125 particles are initialized on a three dimensional simple cubic lattice, all  spins aligned along the positive $z$-direction. Then the constraint to the lattice is removed and the particles are free to move in the entire simulation box. The system has periodic boundary conditions. The external field is set such that dipoles pointing along the negative $z$-direction minimize the energy. To break the symmetry of the initial state and prevent the system of being stuck in an unstable stationary state, the particles are assigned some random momenta and angular velocities that are small in comparison to their values later along the trajectories. The interaction of the particles is modeled via a purely repulsive Weeks-Chandler-Anderson interaction (specified by the corresponding Lennard-Jones parameters $\epsilon$ and $\sigma$) and an exact magnetic dipole-dipole interaction \cite{Landecker1900} using the nearest image convention. Explicitly, the dipole force $F_{ij}$ and torque $G_{ij}$ on particle $j$ due to particle $i$ is given by:
\begin{align}
	\vec{F}_{ij} &= \frac{3\mu_0}{4\pi r^4}\left(\left(\vec{n}\times\vec{m}_i\right)\times\vec{m}_j+\left(\vec{n}\times\vec{m}_j\right)\times\vec{m}_i\right.\nonumber\\
	&\phantom{=}\left. -2\vec{n}\left(\vec{m}_i\cdot\vec{m}_j\right)+5\vec{n}\left(\vec{n}\times\vec{m}_i\right)\cdot\left(\vec{n}\times\vec{m}_j\right) \right)\\
	\vec{G}_{ij} &= \frac{\mu_0}{4\pi r^3}\left(3\left(\vec{m}_i\cdot\vec{n}\right)\vec{m}_j\times\vec{n}+\vec{m}_i\times\vec{m}_j\right)
\end{align}
where $r$ is the distance of the dipoles, $\vec{n}$ is the normalized vector pointing from the position of particle $i$ to $j$, $\mu_0$ is the permeability of free space and $\vec{m}_{i,j}$ are the dipole moments of the particles. The propagation of the system is done using the velocity-Verlet integrator. The particles with a mass $m$ and moment of inertia $I=m\sigma^2/4$ are placed in a box of volume $(6\sigma)^3$. The dipole moment of the individual particles $\vec{m}$ is chosen such that $\epsilon=2\mu_0\left\|\vec{m}\right\|^2/\left(\pi\sigma^3\right)$. As a final remark, it should be mentioned that the timescale $\delta t$ used in \cref{fig:observables,fig:kernelswidthcomparison} is arbitrary because it can be trivially scaled by the particle masses and moments of inertia or the energies and dipole moments. 

As an observable for this transition, we compute the total dipole moment along the $z$-axis (scaled and shifted as explained before). The averaged observable of 1000 trajectories is depicted in \cref{fig:observables} (Dipoles, dash-dotted line). The dipole moment shows a strong overshoot such that the naively determined duration of the transition, as done for the Potts model, is not a good indicator for the relaxation time. To generate an observables that indicates the progress of relaxation, we integrated the modulus of the derivative of this curve. Rescaling and shifting as before yields the dashed red curve in \cref{fig:observables} (Dipoles: Monotonic). We determine the duration of the transition as well as the time at which the extent of the kernel is evaluated from this curve. The corresponding kernel is also shown in \cref{fig:kernelswidthcomparison} and the ratio of the duration of the transition and the extent of the kernel is given in \cref{table:ratiosWidthTransitionTime}. Again, the ratio is of order one.

\section{Anharmonic Spring System}
\begin{figure}
	\centering
	\includegraphics[scale=2]{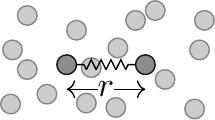}
	\caption{Sketch of anharmonic spring system. Here, all particles interact via a repulsive short range interaction. Additionally, two particles with distance $r$ are coupled via an anharmonic spring.}
	\label{fig:springsketch}
\end{figure}
\begin{figure}
	\centering
	\includegraphics[width=0.89\linewidth]{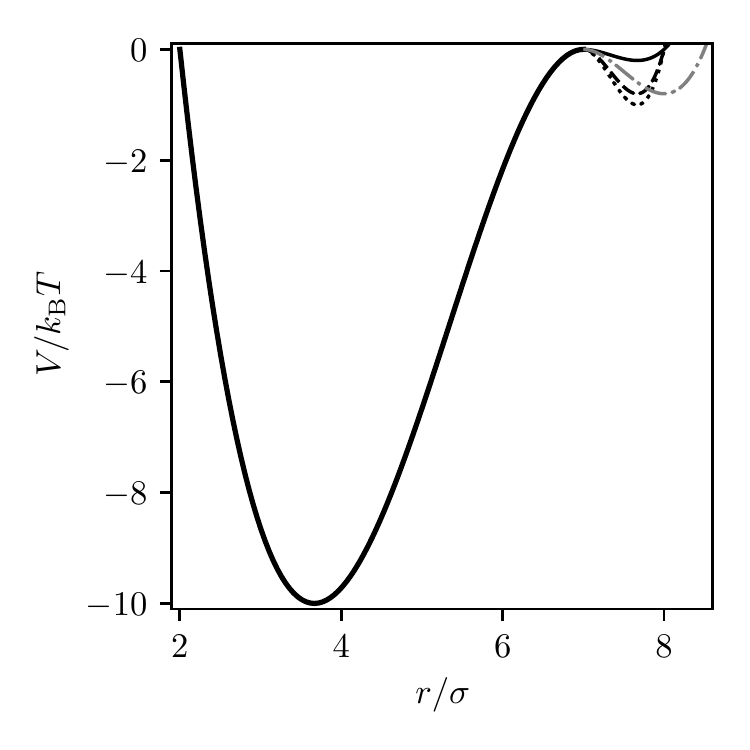}
	\caption{Interaction potentials of anharmonic spring. The interaction potentials are piecewise defined for $0\leq r \leq 7\sigma$ and $7\sigma < r$. Only the long-range part of the interaction potential is varied.}
	\label{fig:springpotentials}
\end{figure}
\begin{figure}
	\centering
	\includegraphics[width=0.99\linewidth]{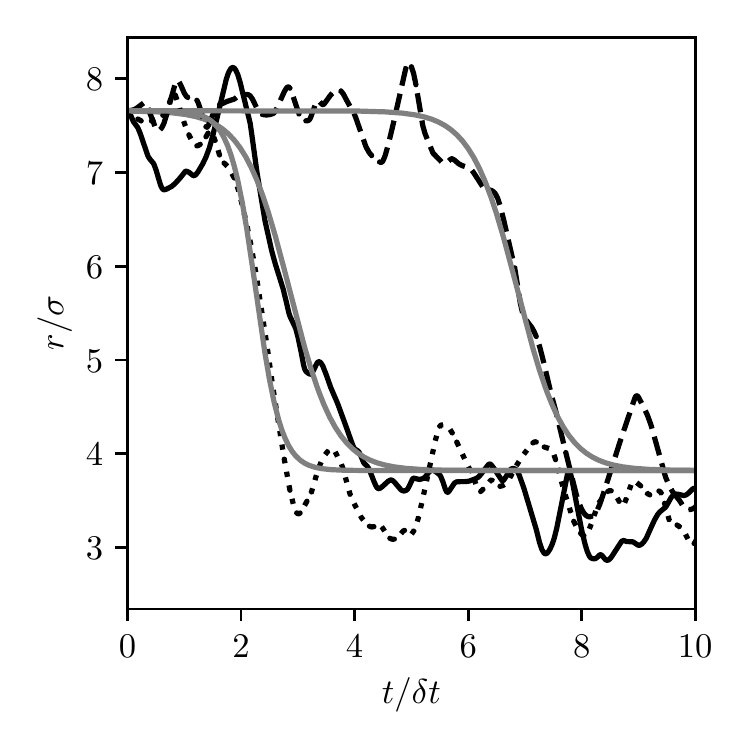}
	\caption{Three exemplary trajectories (black curves) obtained from an interaction potential with a potential depth of $1.0\,k_\text{B}T$ and fits of the sigmoidal form (gray lines).}
	\label{fig:springtrajectories}
\end{figure}
In this section, we test the dependence of the memory kernel on the distribution of transition times $p_T(T)$. We consider a system of particles interacting via the Weeks-Chandler-Anderson potential (without dipoles). Additionally, two particles are coupled by an anharmonic spring with an interaction potential $V(r)$, where $r$ denotes their distance (compare \cref{fig:springsketch}). We pull the coupled particles apart and then observe the relaxation of their distance (which is a ``coarse-grained observable'', because the remaining particles form an effective bath.). A suitable choice of the form of the anharmonicity then allows us to tune $p_T(T)$ independently from $\Delta$.

The interaction potential is defined piecewise by two polynomials of degree three. These two polynomials are chosen such that the interaction potential and its derivative are continuous. The interaction changes from one polynomial to the other in a double zero at $r=7\sigma$ (see \cref{fig:springpotentials}), where a potential barrier is located. Both polynomials are completely determined by the position of their remaining zero and their potential depth. For the short range part, we always set $V(2\sigma)=0$ and the potential depth is $-10k_\text{B}T$. The form of the long range potential is altered in different simulations and takes one of the following forms:\\
In three cases, we choose $V(8\sigma)=0$ and potential depths $\{0.2k_\text{B}T,\,0.8k_\text{B}T,\,1.0k_\text{B}T\}$ corresponding to the black lines in \cref{fig:springpotentials}. The gray line in \cref{fig:springpotentials} has $V(8.5\sigma)=0$ and a potential depth of $0.8k_\text{B}T$.
For each of these four interaction potentials the simulations are done as follows:\\
All particles are placed in the simulation box without any overlap. The distance of the two particles connected via the anharmonic spring is set to the location of the local minimum of $V(r)$ around $r=7.6\sigma$. Then, velocities according to the temperature $T$ are assigned to all particles except the ones connected by the spring. Note, that this is the temperature used to specify other system parameters such as the depths of the spring potential. This is important because the temperature may slightly vary during the simulation due to a systematic increase in the Weeks-Chandler-Anderson interaction energy and a decrease in the spring energy when the spring gets contracted towards its global minimum of the potential $V(r)$. At the beginning, the two particles connected via the spring are kept at fixed positions to ensure that the remaining system is equilibrated before releasing them. Once the particles are released, their distance is tracked as the order parameter of the relaxation process.

\begin{figure}
	\centering
	\includegraphics[width=0.99\linewidth]{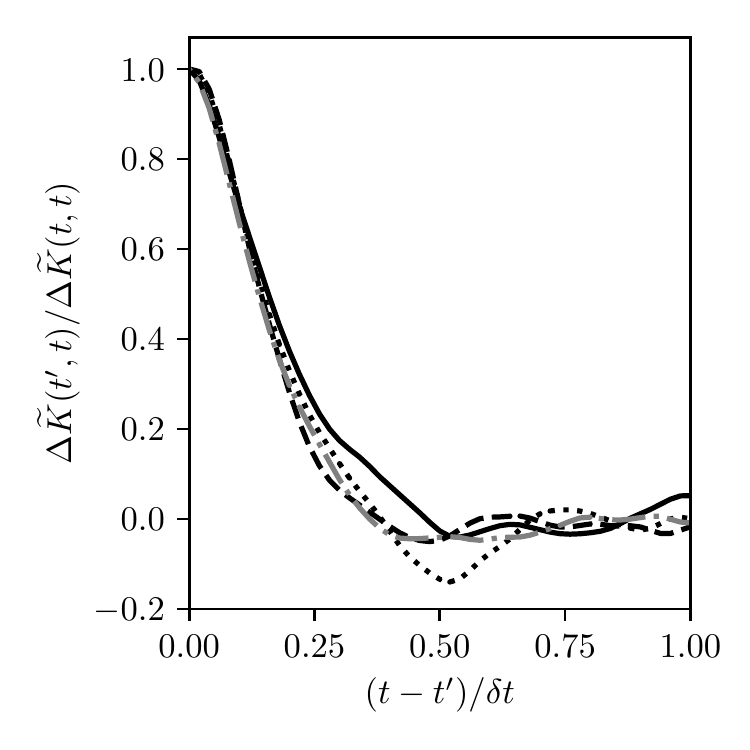}
	\caption{Normalized memory kernels at their respective averaged transition time for the four spring systems, i.e.~for different $p_T(T)$. The line styles for the different spring potentials correspond to the ones used in \cref{fig:springpotentials}.}
	\label{fig:springkernels}
\end{figure}

A simulation box of size $(20\sigma)^3$ containing $4000$ particles is used. The interaction strength of the Weeks-Chandler-Anderson is set to $\epsilon =k_\text{B}T$. As before, the simulation box is assumed to have periodic boundaries and the particles are propagated using the velocity-Verlet integrator. Again, the particle masses affect the dynamics only by rescaling the timescale $\delta t$. For each of the four spring potentials $600$ simulations are carried out.\\
As can be seen in \cref{fig:springtrajectories}, the distance as a function of time shows the desired behavior. These trajectories are well described with the sigmoid defined in \cref{eq:OPdynamics} if one adds an additional offset, scales the transition width with a prefactor and assumes a negative duration of the transition $\Delta \to -\Delta$. (The last point is due to the fact that the distance relaxes from bigger values in the beginning to smaller ones in the end.) Fitting the individual trajectories to such sigmoidal functions where only the induction time $T$ and the duration of the transition $\Delta$ are left as free fit parameters, one obtains a good agreement between the original data and the fits (see gray lines in \cref{fig:springtrajectories}). From these fits, we obtain the averaged induction time $\langle T\rangle$ of the transitions as well as the averaged duration $\langle \Delta\rangle$ of the transitions. By regarding the kernels of the four processes at their respective averaged induction times (see \cref{fig:springkernels}), one can see that they differ only slightly. Especially, the kernel extents vary only weakly (maximum relative deviation $\sim14\%$) whereas the averaged induction times deviates drastically (maximum relative deviation $\sim76\%$). The explicit values are given in \cref{table:springSystem} where also the ratio of kernel extends to the duration of the transitions is given. Again, these ratios are of order one.

\begin{table}[]
	\begin{tabular}{c|c|c|c}
		Spring Potential & $\langle T\rangle/\delta t$ & Kernel Extent$/\delta t$ & Ratio\\
		\hline
		\hline
		$\{0.2\kB T,\,8.0\sigma\}$ & $6.6$ & $0.16$ & $0.33$\\
		$\{0.8\kB T,\,8.0\sigma\}$ & $8.1$ & $0.14$ & $0.28$\\
		$\{1.0\kB T,\,8.0\sigma\}$ & $9.9$ & $0.15$ & $0.31$\\
		$\{0.8\kB T,\,8.5\sigma\}$ & $11.6$ & $0.14$ & $0.26$\\
	\end{tabular}
	\caption{Averaged induction time, kernel extent and ratio of kernel extent to duration of transition for different spring potentials. As before, the spring potential is determined by the depth of the second minimum and the third zero point of the interaction potential.}
	\label{table:springSystem}
\end{table}

\section{Hard-Sphere Nucleation}
As a more complex example, we study the crystal nucleation and growth process in a compressed liquid of hard spheres.
The simulation employs an event driven molecular dynamics (EDMD) algorithm, following mostly the method proposed by Bannermann, due to its robustness regarding overlaps \cite{bannerman:2014}. 
The system consists of 16384 particles and is enclosed by cubic periodic boundary conditions at an initial volume fraction of $\eta_{\text{sim}} = 0.45$, corresponding to the stable liquid phase. The particles' diameter $\sigma$ and mass $\text{m}$ as well as the thermal energy $\text{k}_\text{B} \text{T}$ are chosen as units of the simulation resulting in a time scale of $\delta t = \sqrt{m / (k_B T)} \, \sigma $. 

The particles are initally placed on a fcc-lattice and the amplitude of their starting velocities is determined by the equipartion theorem to be $|\vec{v}|= \sqrt{3} \, \frac{\sigma}{\delta t}$. The directions are chosen at random, under the constraint to keep the center of mass at rest. This inital configuration is equilibrated for about $\text{T}_{\text{equi}}/\delta t = 250$, after which the system has formed a stable liquid.

The system is then rapidly compressed to a volume fraction of $\eta = 0.54$ in about $\text{T}_{\text{comp}}/\delta t = 0.3$. The compression is implemented by rescaling all positions as well as the box length, and afterwards resolving all overlaps by the dynamics of the system.

We monitor the crystalinity of the system in terms of the q6q6-bond-order parameters (\cite{steinhardt:1983}, \cite{tenWolde:1995}).
The local structure around a particle $i$ with $N_b$ neigbours is characterized by the quantity
\begin{equation}
\label{eqn:local_q6}
\bar{q}_{lm}(i) = \frac{1}{N_b} \sum^{N_b (i)}_{j=1} Y_{lm} (\hat{r}_{ij})
\end{equation}
where $Y_{lm}(\hat{r}_{ij})$ are the spherical harmonics evaluated in the direction of the relative position of particles i and j in a given coordinate system.

$\bar{q}_{6m}(i)$ suffices to indicate the local fcc structure of hard-sphere crystals. Based on $\bar{q}_{6m}(i)$  a normalized vector $\vec{q}_{6}(i)$ is defined with elements for $m=-6$ to $m=6$ given by:
\begin{equation}
q_{6m}(i) = \frac{\bar{q}_{6m}(i)}{ \sqrt{\sum_{m'=-6}^{6} |{\bar{q}_{6m'}(i)})}|  }
\end{equation}

As a minimum threshold for the scalar product $\vec{q}_6(i) \cdot \vec{q}_6(j)$ we choose 0.6 to identify a pair of particles $i$ and $j$ as ``orientationally bonded''. To define a solid particle we set the minimum number of bonded neighbours to be $n_B > 8$, similar to e.g. \cite{tenWolde:1995} \cite{schilling:2011}.

As an observable we measure the fraction of solid particles in the system.
A total number of $580$ trajectories are simulated of which $562$ pass through the phase transition within the simulation time. By performing a least square fit using the sigmoid function eq.~\ref{eq:OPdynamics} to the individual solid fraction trajectories and averaging the optimal values, we obtain the duration of the transition as well as the time at which the extent of the kernel is evaluated.

\begin{figure}[]
	\centering
	\includegraphics[width=0.99\linewidth]{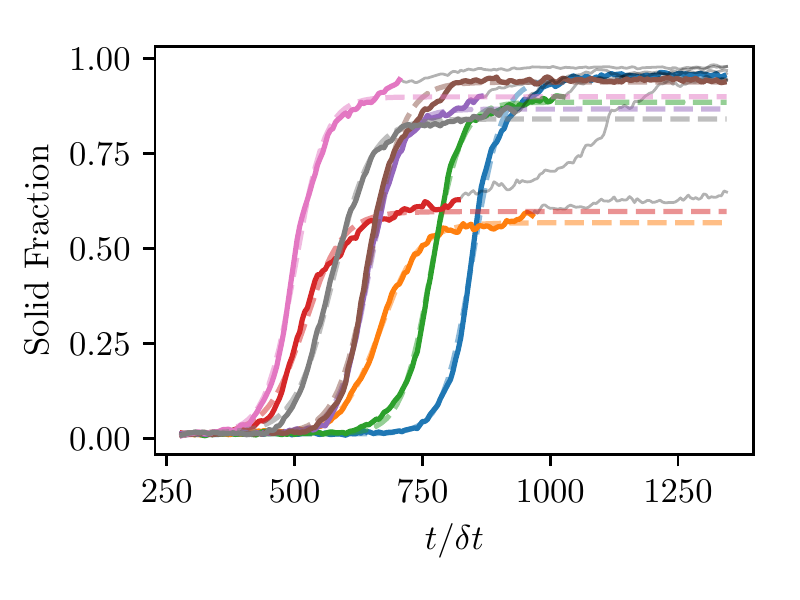}
	\caption{Example trajectories of the solid fraction in the hard-sphere system. The dashed line represents the sigmoidal best fit to the colored data, while the grey data points have been excluded.}
	\label{fig:nucleationtrajectories}
\end{figure}

Fig.~\ref{fig:nucleationtrajectories} shows a subset of the simulated trajectories. Some trajectories reach an intermediate plateau crystallizing in more than one step. In those cases the growing clusters contain defects. For the fit only data points from the first step are used, corresponding to the actual crystal nucleation event, and not to the stage of healing defects. This choice introduces a bias on the transition times, however, the effect on the estimate of $\Delta$ is of about 10\% which seems acceptable for the comparison of magnitudes we carry out here.

\begin{figure}
	\centering
	\includegraphics[width=0.99\linewidth]{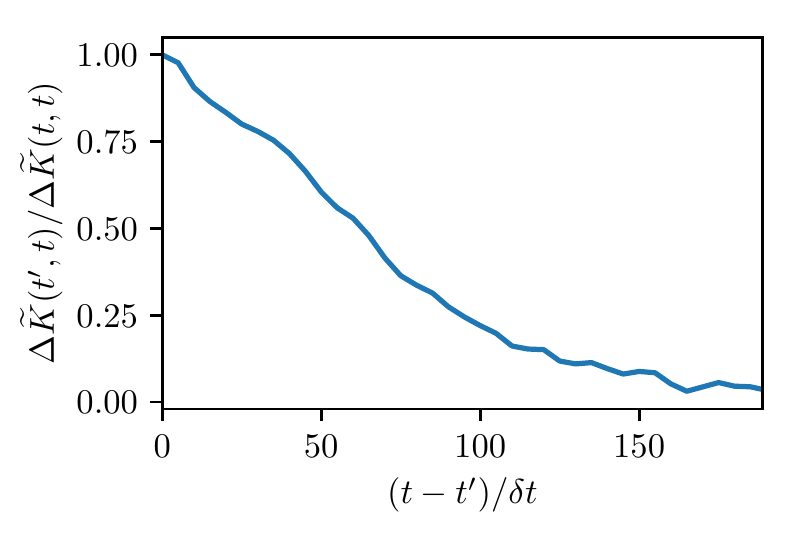}
	\caption{Kernel of crystal nucleation and growth in hard spheres.}
	\label{fig:kernelwidthnucleation}
\end{figure}

For the calculation of the kernel (see Fig \cref{fig:kernelwidthnucleation}) we use all trajectories. The ratio of the extent of the kernel to the duration of the transition is given in \cref{table:ratiosWidthTransitionTime} and is again of similar order of magnitude.


\section{Conclusion}

We have shown in this paper that the memory effects in the order parameter dynamics at a phase-transition can be robustly correlated to the duration of the transition, i.e. the timescale needed for a system to go from the initial phase to the final one. In contrast, the distribution of induction times has no significant impact on the extent of memory. This result was first theoretically predicted and then backed up by simulations of different systems. 

The correlation can be seen as a useful practical tool for anyone interested in modelling the dynamics of a phase transition, who has some phenomenological knowledge (based on experimental observations or numerical data). In fact, one could then use a non-stationary Generalized Langevin Equation of the form \ref{EOM_A} for which the characteristics of the memory kernel would be guided by the phenomenological knowledge one has of the process under study. This is, to our opinion, a novel approach to describe quantitatively the dynamics of phase-transitions, that allows to overcome the inaccuracies of Markovian quasi-equilibrium models by using an exactly derived Generalized Langevin Equation, whose parameters can be chosen using the experimental knowledge one has of the problem.

\section*{Acknowledgements}

This project has been financially supported by the National Research Fund Luxembourg (FNR) within the AFR-PhD programme and by the German Research Association, grant number SCHI 853/6-1. M acknowledges financial support by the DFG via the Collaborative Research Center SFB 1027.


\clearpage
\appendix

\section{Impact of the noise $\xi_{t}$}
\label{xit}

In the main text, we assumed the evolution of an arbitrary order parameter to be of the form 
\begin{align}
\label{evolution_At}
A_{t} =& \frac{1}{1 + e^{-\frac{t-T}{\Delta}}} + \xi_{t} = A^{(0)}_{t} + \xi_{t} 
\end{align}
We assume here 
\begin{equation}
	\left\{
	\begin{tabular}{l}
     $\lla \xi(t) \rra = 0$\\
	 $\lla \xi(t)A^{(0)}(t) \rra = 0$\\
	 $\lla \xi(t)\xi(t') \rra = \epsilon \sqrt{\lla A^{(0)}(t)^{2} \rra\lla A^{(0)}(t')^{2} \rra} \chi\left(\frac{t-t'}{T_{\xi}}\right)$
	\end{tabular}
	\right.
\end{equation}
where $\chi$ is such that $\chi(0)=1$ and decays monotonically, with $\chi(u\gg 1) \simeq 0$. Because we consider a noise whose typical timescale is very small, we will look at the limit $T_{\xi}\to 0$. In this case, we want to show here that the impact of the noise $\xi_{t}$ on the overall memory kernel at first order results into local contributions at $t=t'$ in the memory kernel $K(t',t)$ and a global prefactor for $t\neq t'$. To do this, we perform a first-order expansion of our method where our small parameter is $\lla \xi(t)\xi(t') \rra/ \lla A^{(0)}(t)A^{(0)}(t') \rra$. The shifted and normalized observable $\Delta\tilde{A}_{t}$ reads then 
\begin{align}
	\Delta\tilde{A}_{t} =& \frac{A_{t} - \lla A(t) \rra}{\sqrt{\lla A(t)^{2} \rra}} \nonumber \\
	=& \frac{A^{(0)}_{t} - \lla A^{(0)}(t) \rra - \xi_{t}}{\sqrt{\lla A^{(0)}(t)^{2} \rra + \lla \xi(t)^{2} \rra}} \nonumber \\
	=& \frac{A^{(0)}_{t} - \lla A^{(0)}(t) \rra - \xi_{t}}{\sqrt{\lla A^{(0)}(t)^{2} \rra}\sqrt{1+ \frac{\lla \xi(t)^{2} \rra}{\lla A^{(0)}(t)^{2} \rra}}} \nonumber \\
	=& \frac{A^{(0)}_{t} - \lla A^{(0)}(t) \rra - \xi_{t}}{\sqrt{\lla A^{(0)}(t)^{2} \rra}}\left( 1 - \frac{\lla \xi(t)^{2} \rra}{2\lla A^{(0)}(t)^{2} \rra} \right) 
\end{align}
By defining a normalized version of the noise as $\tilde{\xi}_{t} = \xi_{t/}\sqrt{\lla A^{(0)}(t)^{2} \rra}$, we obtain 
\begin{widetext}
	\begin{align}
	\lla \Delta\tilde{A}(t) \Delta\tilde{A}(t') \rra =& \lla \left(\Delta\tilde{A}^{(0)}_{t} - \tilde{\xi}_{t}\right)\left( 1 - \frac{\lla \tilde{\xi}(t)^{2} \rra}{2} \right)  \left(\Delta\tilde{A}^{(0)}_{t'} - \tilde{\xi}_{t'}\right)\left( 1 - \frac{\lla \tilde{\xi}(t')^{2} \rra}{2} \right) \rra \nonumber \\
	=& \lla \left(\Delta\tilde{A}^{(0)}_{t} - \tilde{\xi}_{t}\right) \left(\Delta\tilde{A}^{(0)}_{t'} - \tilde{\xi}_{t'}\right) \rra \left( 1 - \frac{\lla \tilde{\xi}(t)^{2} \rra + \lla \tilde{\xi}(t')^{2} \rra}{2} \right) \nonumber \\
	=& \left( \lla \Delta\tilde{A}^{(0)}(t) \Delta\tilde{A}^{(0)}(t') \rra + \lla \tilde{\xi}(t) \tilde{\xi}(t') \rra \right)\left( 1 - \frac{\lla \tilde{\xi}(t)^{2} \rra + \lla \tilde{\xi}(t')^{2} \rra}{2} \right) \nonumber \\
	=&  \lla \Delta\tilde{A}^{(0)}(t) \Delta\tilde{A}^{(0)}(t') \rra   \left( 1 - \frac{\lla \tilde{\xi}(t)^{2} \rra + \lla \tilde{\xi}(t')^{2} \rra}{2} \right)  + \lla \tilde{\xi}(t) \tilde{\xi}(t') \rra \nonumber \\
	\Delta \tilde{C}(t',t)=& \Delta \tilde{C}^{(0)}(t',t)   \left( 1 - \epsilon \right)  + \epsilon \chi\left(\frac{t-t'}{T_{\xi}}\right)
	\end{align}
where $ \Delta \tilde{C}^{(0)}(t',t)$ is the contribution obtained by only considering the sigmoidal part of eqn.~(\ref{evolution_At}). We can now compute the first terms $\Delta\tilde{S}_{n}(t',t)$ of the expansion as follows :
\begin{align}
	\Delta\tilde{S}_{0}(t',t) =& \frac{\partial \Delta \tilde{C}(t',t)}{\partial t'} \nonumber \\
	=& \frac{\partial \Delta \tilde{C}^{(0)}(t',t)}{\partial t'} \left(1-\epsilon \right)  - \frac{\epsilon}{T_{\xi}} \chi'\left(\frac{t-t'}{T_{\xi}}\right)  \nonumber \\
	=& \Delta\tilde{S}_{0}^{(0)}(t',t) \left(1-\epsilon \right)   - \frac{\epsilon}{T_{\xi}} \chi'\left(\frac{t-t'}{T_{\xi}}\right) 
\end{align}
where $\Delta\tilde{S}_{n}^{(0)}$ is the contribution to $\Delta\tilde{S}_{n}$ that would be obtained without the noise $\xi_{t}$. We go on to the next order : 
\begin{align}
\label{DS1}
\Delta\tilde{S}_{1}(t',t) =& \int_{t'}^{t} d\tau \Delta\tilde{S}_{0}(t',\tau)\Delta\tilde{S}_{0}(\tau,t) \nonumber \\ 
=& \left(1-\epsilon\right)^{2}\int_{t'}^{t} d\tau \Delta\tilde{S}_{0}^{(0)}(t',\tau) \Delta\tilde{S}_{0}^{(0)}(\tau,t) +\frac{\epsilon^{2}}{T_{\xi}^{2}} \int_{t'}^{t} d\tau \chi'\left(\frac{t-\tau}{T_{\xi}}\right) \chi'\left(\frac{\tau-t'}{T_{\xi}}\right) \nonumber \\ 
&-\frac{\epsilon\left(1-\epsilon\right)}{T_{\xi}}\int_{t'}^{t} d\tau \Delta\tilde{S}_{0}^{(0)}(t',\tau) \chi'\left(\frac{t-\tau}{T_{\xi}}\right) -\frac{\epsilon\left(1-\epsilon\right)}{T_{\xi}}\int_{t'}^{t} d\tau \Delta\tilde{S}_{0}^{(0)}(\tau,t) \chi'\left(\frac{\tau-t'}{T_{\xi}}\right) \nonumber \\ 
=& \left(1-\epsilon\right)^{2} \Delta\tilde{S}_{1}^{(0)}(t',t) +\frac{\epsilon^{2}}{T_{\xi}} \int_{0}^{\frac{t-t'}{T_{\xi}}} du \chi'\left(\frac{t-t'}{T_{\xi}} - u\right) \chi'\left(u\right) \nonumber \\ 
&-\epsilon\left(1-\epsilon\right) \int_{0}^{\frac{t-t'}{T_{\xi}}} d\tau \Delta\tilde{S}_{0}^{(0)}(t',t-uT_{\xi}) \chi'\left(u\right) -\epsilon\left(1-\epsilon\right)\int_{0}^{\frac{t-t'}{T_{\xi}}} d\tau \Delta\tilde{S}_{0}^{(0)}(t' + uT_{\xi},t) \chi'\left(u \right) \nonumber \\  
=& \left(1-\epsilon\right)^{2} \Delta\tilde{S}_{1}^{(0)}(t',t) +\frac{\epsilon^{2}}{T_{\xi}} \left[\chi'(0)\chi\left(\frac{t-t'}{T_{\xi}}\right) - \chi'\left(\frac{t-t'}{T_{\xi}}\right) + \int_{0}^{\frac{t-t'}{T_{\xi}}} du \chi'\left(\frac{t-t'}{T_{\xi}} - u\right) \chi\left(u\right) \right] \nonumber \\ 
&-\epsilon\left(1-\epsilon\right) \left[\left(\Delta\tilde{S}_{0}^{(0)}(t',t') + \Delta\tilde{S}_{0}^{(0)}(t,t)\right) \chi\left(\frac{t-t'}{T_{\xi}}\right) -  2\Delta\tilde{S}_{0}^{(0)}(t',t) \right. \nonumber \\
&\ \ \ \ \ \ \ \ \ \ \ \ \ \ \left.+ T_{\xi} \int_{0}^{\frac{t-t'}{T_{\xi}}} du \left( \frac {\partial\Delta\tilde{S}_{0}^{(0)}}{\partial t'}(t' + uT_{\xi},t) - \frac {\partial\Delta\tilde{S}_{0}^{(0)}}{\partial t}(t',t-uT_{\xi}) \right) \chi\left(u\right)  \right] 
\end{align}
At first order in $\epsilon$ we get 
\begin{align}
\Delta\tilde{S}_{1}(t',t) =  \left(1-2\epsilon\right) \Delta\tilde{S}_{1}^{(0)}(t',t) -\epsilon &\left[\left(\Delta\tilde{S}_{0}^{(0)}(t',t') + \Delta\tilde{S}_{0}^{(0)}(t,t)\right) \chi\left(\frac{t-t'}{T_{\xi}}\right) -  2\Delta\tilde{S}_{0}^{(0)}(t',t) \right. \nonumber \\
&\ \ \ \ \ \  \left.+ T_{\xi}  \int_{0}^{\frac{t-t'}{T_{\xi}}} du \left( \frac {\partial\Delta\tilde{S}_{0}^{(0)}}{\partial t'}(t' + uT_{\xi},t) - \frac {\partial\Delta\tilde{S}_{0}^{(0)}}{\partial t}(t',t-uT_{\xi}) \right) \chi\left(u\right)  \right] \nonumber \\
 =  \left(1-2\epsilon\right) \Delta\tilde{S}_{1}^{(0)}(t',t) -\epsilon &\left[\left(\Delta\tilde{S}_{0}^{(0)}(t',t') + \Delta\tilde{S}_{0}^{(0)}(t,t)\right) \chi\left(\frac{t-t'}{T_{\xi}}\right) -  2\Delta\tilde{S}_{0}^{(0)}(t',t) \right. \nonumber \\
 &\ \ \ \ \ \  \left.+ \int_{0}^{t-t'} d\tau \left( \frac {\partial\Delta\tilde{S}_{0}^{(0)}}{\partial t'}(t' + \tau,t) - \frac {\partial\Delta\tilde{S}_{0}^{(0)}}{\partial t}(t',t-\tau) \right) \chi\left(\frac{\tau}{T_{\xi}}\right)  \right] 
\end{align}
In the limit $T_{\xi}\to 0$, the last term vanishes, and we end up only with 
\begin{align}
\Delta\tilde{S}_{1}(t',t) =  \left(1-2\epsilon\right) \Delta\tilde{S}_{1}^{(0)}(t',t) +\epsilon   2\Delta\tilde{S}_{0}^{(0)}(t',t) - \epsilon \left(\Delta\tilde{S}_{0}^{(0)}(t',t') + \Delta\tilde{S}_{0}^{(0)}(t,t)\right) \chi\left(\frac{t-t'}{T_{\xi}}\right)
\end{align}
which can also be written as
\begin{equation}
	\Delta\tilde{S}_{1}(t',t) = \left\{ 
	\begin{tabular}{ll}
		 $\left(1-2\epsilon\right) \Delta\tilde{S}_{1}^{(0)}(t',t) +\epsilon   2\Delta\tilde{S}_{0}^{(0)}(t',t)$ & if $|t-t'| > 0$ \\
		 $\left(1-2\epsilon\right) \Delta\tilde{S}_{1}^{(0)}(t,t) $ & if $t=t'$ \\
	\end{tabular}
	\right.
\end{equation}
To compute the next order, we can then use the expression for $\Delta\tilde{S}_{1}(t',t)$ for $|t-t'| > 0$ because the amplitude of $\chi$ remains finite and thus any convolution product with it will vanish in the limit $T_{\xi}\to 0$. We therefore infer at any order $n>0$ for $|t-t'|>0$ the identity 
\begin{equation}
	\Delta\tilde{S}_{n}(t',t) =   (1-(n+1)\epsilon) \Delta\tilde{S}_{n}^{(0)}(t',t) + (n+1) \epsilon   \Delta\tilde{S}_{n-1}^{(0)}(t',t) 
\end{equation}
We prove this by induction, following the same steps as the ones we used for $\Delta\tilde{S}_{1}$ (see \ref{DS1}), again in the limit $T_{\xi}\to 0$ and for $|t-t'|>T_{\xi}$ :
\begin{align}
\Delta\tilde{S}_{n+1}(t',t) =& \int_{t'}^{t} d\tau \Delta\tilde{S}_{n}(t',\tau)\Delta\tilde{S}_{0}(\tau,t) \nonumber \\ 
=& \left(1-(n+1)\epsilon\right)\left(1-\epsilon\right)\int_{t'}^{t} d\tau \Delta\tilde{S}_{n}^{(0)}(t',\tau) \Delta\tilde{S}_{0}^{(0)}(\tau,t) +\frac{\left(n+1\right)\epsilon^{2}}{T_{\xi}} \int_{t'}^{t} d\tau \Delta\tilde{S}_{n-1}^{(0)}(t',\tau) \chi'\left(\frac{t-\tau}{T_{\xi}}\right)  \nonumber \\ 
&-\frac{\epsilon\left(1-(n+1)\epsilon\right)}{T_{\xi}}\int_{t'}^{t} d\tau \Delta\tilde{S}_{n}^{(0)}(t',\tau) \chi'\left(\frac{t-\tau}{T_{\xi}}\right) +(n+1)\epsilon\left(1-\epsilon\right)\int_{t'}^{t} d\tau \Delta\tilde{S}_{n-1}^{(0)}(t',\tau) \Delta\tilde{S}_{0}^{(0)}(\tau,t)  \nonumber \\ 
=& \left(1-(n+2)\epsilon\right) \Delta\tilde{S}_{n+1}^{(0)}(t',\tau) + \epsilon \Delta\tilde{S}_{n}^{(0)}(t',t)  +(n+1)\epsilon \Delta\tilde{S}_{n}^{(0)}(t',\tau)   \nonumber \\ 
=& \left(1-(n+2)\epsilon\right) \Delta\tilde{S}_{n+1}^{(0)}(t',\tau)   +(n+2)\epsilon \Delta\tilde{S}_{n}^{(0)}(t',\tau)  
\end{align}
\end{widetext}
This proves the iterative relation. Now, we compute the expansion $\Delta\tilde{\mathcal{S}}(t',t) = \sum_{n=0}^{\infty}\Delta\tilde{\mathcal{S}}_{n}(t',t)$ as follows : 
\begin{align}
	\Delta\tilde{\mathcal{S}} =& \sum_{n=0}^{\infty}\Delta\tilde{\mathcal{S}}_{n} \nonumber \\
	=& \Delta\tilde{\mathcal{S}}_{0} + \sum_{n=1}^{\infty}\Delta\tilde{\mathcal{S}}_{n} \nonumber \\
	=& (1-\epsilon)\Delta\tilde{\mathcal{S}}_{0}^{(0)} + \sum_{n=1}^{\infty}\Delta\tilde{\mathcal{S}}_{n}^{(0)} \nonumber \\
	&- \epsilon \left( \sum_{n=1}^{\infty}(n+1) \Delta\tilde{\mathcal{S}}_{n}^{(0)} - \sum_{n=1}^{\infty}(n+1) \Delta\tilde{\mathcal{S}}_{n-1}^{(0)} \right)\nonumber \\
	=& \Delta\tilde{\mathcal{S}}^{(0)} -\epsilon \Delta\tilde{\mathcal{S}}_{0}^{(0)} \nonumber \\
	&- \epsilon \left( \sum_{n=1}^{\infty}(n+1) \Delta\tilde{\mathcal{S}}_{n}^{(0)} - \sum_{n=0}^{\infty}(n+2) \Delta\tilde{\mathcal{S}}_{n}^{(0)} \right)\nonumber \\
	=& \Delta\tilde{\mathcal{S}}^{(0)} -\epsilon \Delta\tilde{\mathcal{S}}_{0}^{(0)} - \epsilon \left( -\sum_{n=1}^{\infty} \Delta\tilde{\mathcal{S}}_{n}^{(0)} - 2\Delta\tilde{\mathcal{S}}_{0}^{(0)} \right)\nonumber \\
	=& \Delta\tilde{\mathcal{S}}^{(0)}  + \epsilon \sum_{n=0}^{\infty} \Delta\tilde{\mathcal{S}}_{n}^{(0)} \nonumber \\
	\Delta\tilde{\mathcal{S}} =& (1+\epsilon) \Delta\tilde{\mathcal{S}}^{(0)} 
\end{align}
We thus prove that apart from the points $t=t'$, considering the autocorrelation of the noise term $\xi_{t}$ only results into a factor $1+\epsilon$ for the anti-derivative of the memory kernel, which does not impact the evaluation of the width of the memory kernel. 

\newpage
\section{The case $\Delta \to 0$}
\label{Delta0}
Let us now check this by explicitly computing the expansion $\Delta\tilde{\mathcal{S}}(t',t) = \sum_{n=0}^{\infty}\Delta\tilde{\mathcal{S}}_{n}(t',t)$. Note that because we work with the shifted and normalized ttACF $\Delta\tilde{C}$, we have $\Delta\tilde{\omega}= 0$, and $\Delta\tilde{\mathcal{S}} = -\Delta\tilde{\mathcal{J}}$. We define
\begin{align}
\left\{
\begin{tabular}{l}
$f_{T}(t) := \int_{0}^{t} d\tau \ p_{T}(\tau)$  \\
$g_{T}(t) := f_{T}(t)/\left(1-f_{T}(t)\right)$ \\
$h_{T}(t) := f_{T}(t)\left(1-f_{T}(t)\right)$  
\end{tabular}
\right.
\end{align}

We start by computing $\Delta\tilde{\mhc{S}}_{0}$, assuming $t'\leq t$:
\begin{align}
\Delta\tilde{\mhc{S}}_{0}(t',t) =& \frac{\partial \Delta \tilde{C}(t',t)}{\partial t'} \nonumber \\
=& \frac{f'_{T}(t')\left(1-f_{T}(t)\right)}{\sqrt{h_{T}(t)h_{T}(t')}} - \frac{1}{2} \frac{f_{T}(t')\left(1-f_{T}(t)\right)}{\sqrt{h_{T}(t)h_{T}(t')}} \frac{h'_{T}(t')}{h_{T}(t')}\nonumber \\
=& \frac{\left(1-f_{T}(t)\right)}{\sqrt{h_{T}(t)h_{T}(t')}} \left(f'_{T}(t')  - \frac{1}{2} \frac{h'_{T}(t')}{1 - f_{T}(t')} \right)\nonumber \\
=& \frac{f'_{T}(t')\left(1-f_{T}(t)\right)}{\sqrt{h_{T}(t)h_{T}(t')}} \left(1  - \frac{1 - 2f_{T}(t')}{2\left( 1 -f_{T}(t')\right)} \right)\nonumber \\
=& \frac{1}{2}\frac{f'_{T}(t')}{\sqrt{h_{T}(t)h_{T}(t')}} \frac{1-f_{T}(t)}{1 -f_{T}(t')} \nonumber \\
=& \frac{1}{2}\frac{p_{T}(t')}{\left(1 - f_{T}(t')\right)\sqrt{h_{T}(t')}} \sqrt{\frac{1}{g_{T}(t)}} \nonumber \\
=& \frac{1}{2}\frac{p_{T}(t')}{\left(1 - f_{T}(t')\right)\sqrt{g_{T}(t')h_{T}(t')}} \sqrt{\frac{g_{T}(t')}{g_{T}(t)}} \nonumber \\
=& \frac{1}{2}\frac{p_{T}(t')}{\left(1 - f_{T}(t')\right)\sqrt{g_{T}(t')h_{T}(t')}} \sqrt{\frac{g_{T}(t')}{g_{T}(t)}} \nonumber \\
\Delta\tilde{\mhc{S}}_{0}(t',t) =& \frac{p_{T}(t')}{2h_{T}(t')} \sqrt{\frac{g_{T}(t')}{g_{T}(t)}}
\end{align}
We continue with the next order, i.e. 
\begin{align}
\Delta\tilde{\mhc{S}}_{1}(t',t) =& \int_{t'}^{t} d\tau S_{0}(t',\tau) S_{0}(\tau,t) \nonumber \nonumber \\
=& \frac{1}{4} \frac{p_{T}(t')}{h_{T}(t')} \sqrt{\frac{g_{T}(t')}{g_{T}(t)}} \int_{t'}^{t} d\tau \frac{p_{T}(\tau)}{h_{T}(\tau)} \nonumber \\
=& \frac{1}{2} S_{0}(t',t) \int_{t'}^{t} d\tau \frac{f'_{T}(\tau)}{f_{T}(\tau)\left(1 - f_{T}(\tau) \right)} \nonumber \\
=& \frac{1}{2} S_{0}(t',t) \ln \left( \frac{g_{T}(t)}{g_{T}(t')} \right)\nonumber \\
\Delta\tilde{\mhc{S}}_{1}(t',t) =& S_{0}(t',t) \ln \left(\sqrt{\frac{g_{T}(t)}{g_{T}(t')}} \right)
\end{align}
We can similarly compute the next orders (see appendix) and find:
\begin{align}
\Delta\tilde{\mhc{S}}_{2}(t',t) =& \frac{1}{2} \Delta\tilde{\mhc{S}}_{0}(t',t) \ln\left(\sqrt{ \frac{g_{T}(t)}{g_{T}(t')}} \right)^{2}\\
\Delta\tilde{\mhc{S}}_{3}(t',t) =& \frac{1}{6} \Delta\tilde{\mhc{S}}_{0}(t',t) \ln\left(\sqrt{ \frac{g_{T}(t)}{g_{T}(t')}} \right)^{3}
\end{align}
From these first orders, we infer 
\begin{equation}
\Delta\tilde{\mhc{S}}_{n}(t',t) = \frac{1}{n!} \Delta\tilde{\mhc{S}}_{0}(t',t) \ln\left(\sqrt{ \frac{g_{T}(t)}{g_{T}(t')}} \right)^{n}
\end{equation}
which implies
\begin{align}
\Delta\tilde{\mhc{J}}(t',t) =& - \sum_{n=0}^{\infty}  \Delta\tilde{\mhc{S}}_{n}(t',t) \nonumber \\
=& - \Delta\tilde{\mhc{S}}_{0}(t',t) \sum_{n=0}^{\infty} \frac{1}{n!}  \ln\left(\sqrt{ \frac{g_{T}(t)}{g_{T}(t')}} \right)^{n}\nonumber \\
=& - \Delta\tilde{\mhc{S}}_{0}(t',t) \exp\left( \ln\left(\sqrt{ \frac{g_{T}(t)}{g_{T}(t')}} \right) \right)\nonumber \\
=& - \Delta\tilde{\mhc{S}}_{0}(t',t) \sqrt{ \frac{g_{T}(t)}{g_{T}(t')}}\nonumber \\
\Delta\tilde{\mhc{J}}(t',t) =& - \frac{p_{T}(t')}{2h_{T}(t')}
\end{align}
We can do the same analysis for $t\leq t'$ and find
\begin{equation}
\Delta\tilde{\mhc{J}}(t',t) = \frac{p_{T}(t')}{2h_{T}(t')}
\end{equation}
Thus, for a fixed $t'$, $\Delta\tilde{\mhc{J}}$ is a step function whose discontinuity is located at $t=t'$ and that jumps from $\frac{p_{T}(t')}{2h_{T}(t')}$ to $-\frac{p_{T}(t')}{2h_{T}(t')}$. Because $\Delta\tilde{K}(t',t)$ is the derivative of $\Delta\tilde{J}$ with respect to t, it is therefore a Dirac function whose peak is located at $t'=t$. Note that this result was obtained without any assumption on the distribution $p(T)$, we thus conclude that the extent of the memory kernel of the averaged phase transition dynamics vanishes when the timescale of the individual transition tends to $0$. This is a universal “kinetic” feature that does not depend on the microscopic details which determine the distribution $p(T)$.

\begin{figure}
	\begin{center}
		\includegraphics[width=0.99\linewidth]{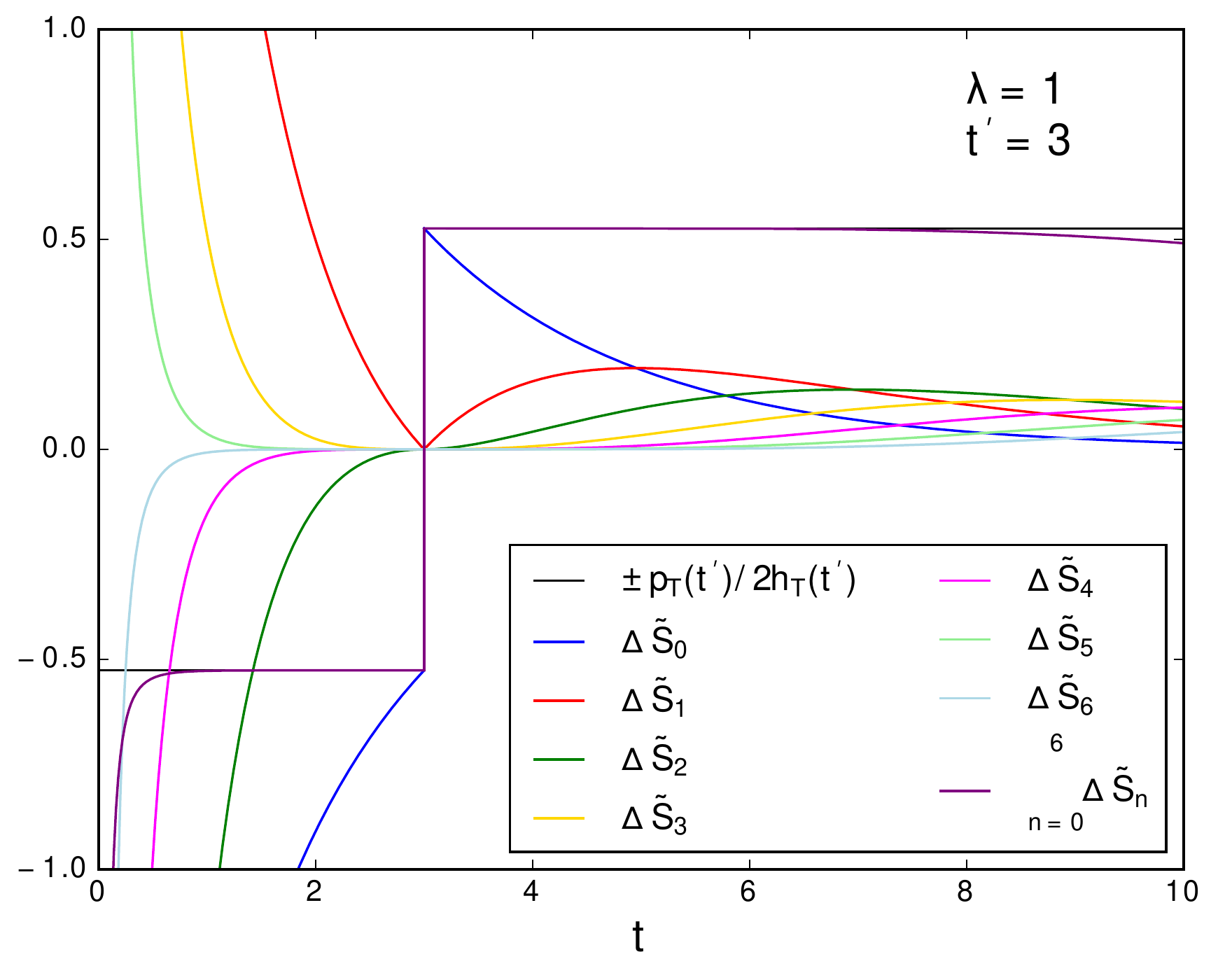}
	\end{center}
	\caption{First terms $\Delta\tilde{S}_{n}(t',t)$ for $n\leq 6$ for a fixed $t'=1$ as a function of $t$, together with their sum and its limit $p_{T}(t')/2h_{T}(t')$ in the case of an exponential distribution $p_{T}(T)=e^{-\lambda T}$.}
	\label{Sn_jump}
\end{figure}

We show in figure \ref{Sn_jump} the first terms $\Delta\tilde{S}_{n}(t',t)$ of the expansion and their sum. We clearly see how they add up to reach the limit of the step function, which yields in turn a Dirac-peaked memory kernel.

\end{document}